\documentclass[10pt]{article}

\usepackage{amsmath}
\usepackage{amssymb}
\usepackage{graphicx}
\usepackage{cite}

\usepackage{color}
\usepackage{setspace}

\usepackage[labelfont=bf,labelsep=period,justification=raggedright]{caption}

\makeatletter
\renewcommand{\@biblabel}[1]{\quad#1.}
\makeatother

\date{}

\begin{document}

\begin{flushleft}
{\Large
\textbf{Emergence of small-world anatomical networks in self-organizing clustered neuronal cultures}
}
\\
Daniel de Santos-Sierra$^{1,\ast}$,
Irene Sendi\~na-Nadal$^{2,1}$,
Inmaculada Leyva$^{2,1}$,
Juan A. Almendral$^{2,1}$,
Sarit Anava$^{3}$,
Amir Ayali$^{3}$,
David Papo$^1$,
Stefano Boccaletti$^{4,5}$
\\
\bf{1} Center for Biomedical Technology, Universidad Polit\'ecnica de Madrid, 28223 Pozuelo de Alarc\'on, Madrid, Spain
\\
\bf{2} Complex Systems Group, Universidad  Rey Juan Carlos, 28933 M\'ostoles, Madrid, Spain
\\
\bf{3} Department of Zoology, Tel-Aviv University, 69978 Tel Aviv, Israel
\\
\bf{4} CNR-Institute of Complex Systems, Via Madonna del Piano, 10, 50019 Sesto Fiorentino, Florence, Italy
\\
\bf{5} INFN Sezione di Firenze, via Sansone, 1-I-50019 Sesto Fiorentino, Italy\\
$\ast$ E-mail: daniel.desantos@ctb.upm.es
\end{flushleft}

\section*{Abstract}


{\it In vitro} primary cultures of dissociated invertebrate neurons from
locust ganglia are used to experimentally investigate the morphological evolution of
assemblies of living neurons, as they self-organize from collections
of separated cells into elaborated, clustered, networks.
At all the different stages of the culture's development,
identification of neurons' and neurites' location by means of a dedicated software
allows to ultimately extract an adjacency matrix from each image of the culture.
In turn, a  systematic statistical analysis of a group of topological
observables grants us the possibility of quantifying and tracking the progression of the main
network's characteristics during the self-organization process of the culture.
Our results point to the existence of a particular state corresponding to a {\it small-world}
network configuration, in which several relevant graph's micro- and meso-scale properties
emerge.
Finally, we identify the main physical processes ruling the culture's
morphological transformations, and embed them into a simplified growth model qualitatively reproducing the
overall set of experimental observations.

\section*{Introduction}

The issue of why and how an assembly of isolated (cultured) neurons self-organizes to
form a complex neural network is a fundamental problem
\cite{Marom2002,vanPelt2005,PRSoriano}.
Despite their more limited, and yet laboratory-controllable, repertoire of
responses \cite{Marom2002,Ayali2004}, the understanding of such cultures' organization is, indeed, a basis for
the comprehension of the mechanisms involved in their {\it in vivo} counterparts,
and provide a useful framework for the investigation of
neuronal network development in real biological systems \cite{PRSoriano}.

Some previous studies highlighted the fact that the structuring of a neuronal cultured
network before the attainment of its mature state is not random, being instead governed and
characterized by processes eventually leading  to configurations which are comparable to many other real complex
networks \cite{PRBocca}. In particular, networking neurons
simultaneously feature a high overall clustering and a relatively short path-length between any pair of
them \cite{PREShefi}.
Such configurations, which in graph theory are termed {\it  small-world} \cite{WS98}, are ubiquitously found in real-world networking
systems. Small-world structures have been shown to enhance the
system's overall efficiency \cite{Latora2003,Achard2007}, while concurrently warranting
a good balance between two apparently antagonistic tendencies for
segregation and integration in structuring processes, needed
for the network's parallel, and yet synthetic performance \cite{PRLRad}.

In this paper, we experimentally investigate the self-organization
into a network of an {\it in vitro} culture of neurons during the course of
development, and explore the changes of the main topological features characterizing the anatomical
connectivity between neurons during the associated network's growth.
To that purpose, dissociated and randomly seeded neurons are initially prepared, and
the spontaneous and self-organized formation of connections is tracked up to their assembling
into a two dimensional clustered network.

Most existing studies in neuronal cultures restricted their attention
to functional networks (statistical dependence between nodes
activities) and not to the physical connections
supporting the functionality of the network \cite{Cossart2011}. The reason behind this
drawback is that the majority of investigations focused on excessively
dense cultures, hindering the observation of their fine scale
structural connectivity. Although there are studies striving to indirectly
infer the underlying anatomical connectivity from the functional
network, it has been shown that strong functional correlations may exist 
with no direct physical connection \cite{Honey2009}. Only few studies dealt with the
physical  wiring circuitry. However, on the one hand, only small networks were considered; on the
other hand, how the network state evolves during the course of the
maturation process has not been investigated \cite{PREShefi}.

Here, instead, we focus on intermediate neurons' densities, and provide a full tracking of
the most relevant topological features emerging during the culture's evolution.
In particular, we show experimentally that {\it in vitro} neuronal
networks tend to develop from a random network state toward a
particular networking state, corresponding to a {\it small-world}
configuration, in which several relevant graph's micro- and meso-scale
properties emerge. Our approach also unveils the main physical processes underlying the culture's
morphological transformation, and allows using such information for
devising a proper growth model, qualitatively reproducing the set of our experimental evidence.

Together with confirming several results of previous works on functional connectivity \cite{Downes2012}, or on
morphological structuring at a specific stage of the cultures'
evolution \cite{PREShefi}, we offer a systematic characterization of several topological network's measures from the very initial until the final state of the culture.
Such a {\it longitudinal} study of the network structure highlights as
yet unknown self-organization properties of cultured neural networks,
such as {\it i}) a large increase in both local and global network's efficiency
associated to the emergence of the small-world configuration, and {\it ii}) the setting of assortative degree-degree
correlation features.

\section*{Experimental set-up}
\subsection*{Neuronal cultures and network growth}
\begin{figure}[h!]
  \centering
  \includegraphics[width=0.7\textwidth]{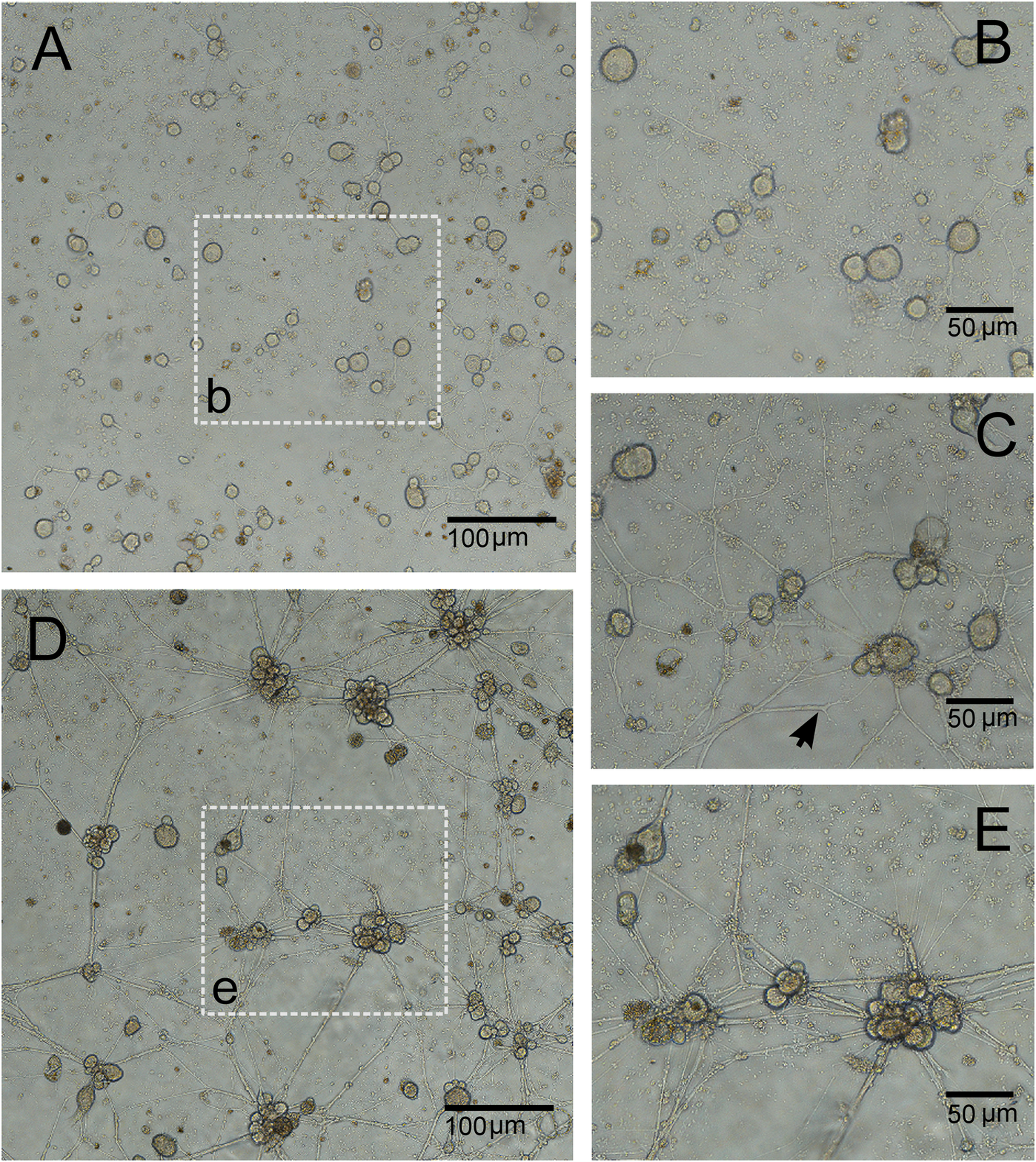}
  \caption{{\bf Culture development of locust frontal ganglion
    neurons into clustered networks}. (A) After 3 DIV, completely
    dissociated neurons had already started growing neuronal processes with
    continuous branching. The area outlined in (b) is enlarged
    in B. (C) Same area as in (B) but at 6 DIV. At this stage, neurons
    and small clusters of neurons are already densely connected and form a
    complex network. At the same stage, branched neurites (pointed by the black arrow) that failed to
    contact neighboring neurons start to retract.  (D) Migration of neurons due to the tension
    along neurites leads to the formation of large neuronal clusters
    and of thicker bundles of neurites. For a better visualization, the area outlined in (e) is enlarged
    in E.}
  \label{fig1}
\end{figure}

In this paper, we report on six cultured networks, which were grown from independent initial sets of dissociated neurons
extracted from the frontal ganglion of adult locusts of the {\it Schistocerca gregaria}
species. In all cases, a same protocol was used, involving animals that were daily fed with organic wheat grass and
maintained under a 12:12 h light:dark cycle from their fifth nymph growth to
their early adult stage of development.  At this latter stage, we
followed the dissection and culturing protocol thoroughly described in
\cite{Anava2013}. In brief, the frontal
ganglia were dissected from anesthetized animals, and enzymatically
treated to soften the sheath. Ganglia were then forced to pass through
the tip of a $200$ $\mu$l pipette to mechanically dissociate the
neurons. The resulting suspension of neuronal somata was plated on
Concanavalin A pre-coated circular area ($r\sim 5$~mm) of a Petri dish
where it was left for $2$ h to allow adhesion of neurons at random positions of the
substrate. After plating, $2$ ml culture medium (Leibovitz L-15)
enriched with 5\% locust hemolymph was added. Cultures were then maintained in darkness under
controlled temperature ($29 ^{\circ}$C) and humidity
($70\%$).

The density at which cultures are seeded determines the maturation rate
and the spatial organization at the mature state
\cite{Shefi2002,Segev2003}. For the purpose of this work, aimed at
studying the network evolution into a clustered network, 6 dense cultures of
12 ganglia each ($\sim 1,200$ neurons)  were used and monitored during
18 days {\it in vitro} (DIV). During the entire experiment, the
culture medium was not changed.

High-resolution and large scale images of the whole culture
were acquired daily using a charge coupled device camera (DS-Fi1,
Nikon) mounted on a phase contrast microscope (Eclipse Ti-S, Nikon), with automated control
of a motorized XYZ stage (H117 ProScan, Prior Scientific).

A typically observed growth evolution is shown in Fig.~\ref{fig1} (restricted to just a
small part of the whole culture) between 3 and 12 DIV.
Neurons ranging from $10$ to $50$ $\mu$m in
size are initially randomly anchored to a two dimensional substrate, while after 3
DIV (Fig.~\ref{fig1}A and B) many cells already start growing neuronal
processes (neurites) trying to target neighboring cells. During
this growth process, neurites also split and reach other processes
forming loops of neurites up to 6 DIV, when the maximum stage of
network development takes place (Fig.~\ref{fig1}C). At this point,
the growth rate decreases and a different mechanism starts shaping the
network: tension is generated
along the neurites as they stretch between neurons or
bifurcation points to form straight segments \cite{Anava2009}.

The latter process favors neuron migration, giving rise to clusters of
neurons, and the fusion of parallel neurites into thicker bundles
together with the retraction of those branches which did not target
any neuron (see black arrow in Fig.~\ref{fig1}C). The resulting
network topology shown in Fig.~\ref{fig1}D after 12 DIV (and in the enlarged
area in Fig.~\ref{fig1}E) is characterized by a random distribution of
few clusters of aggregated neurons linked by thick nerve-like
bundles.

\subsection*{Anatomical graph extraction and complex network statistics}

\begin{figure}
  \centering
\includegraphics[width=\textwidth]{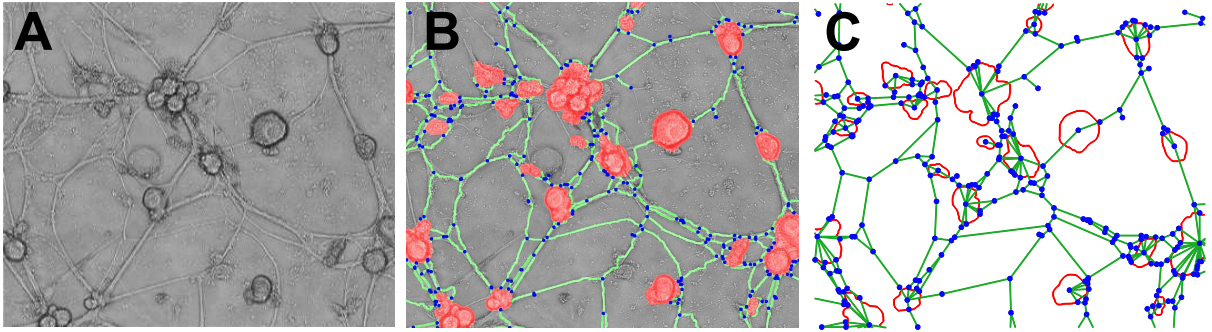}
  \caption{{\bf Extraction of the adjacency matrix defining the neural
    network connectivity}. (A) Image cut taken from a 6 DIV culture and
    (B) the layer on top showing the identification of neurons and
    clusters of neurons (red), neurites connecting them (green) and
    neurite branching points (blue). (C) Mapping of the neuronal
    network into a graph where blue dots represent the nodes and
    green lines the links of the graph.
  }
  \label{fig2}
\end{figure}

Our experiments consistently show that cultures self-organize from random scattered distributions of
bare neurons into spatial networks of interconnected clusters of neurons (compare Fig.\ref{fig1}A and Fig.\ref{fig1}E).

In order to properly quantify the topological and spatial changes of the
anatomical neuronal network as cultures approach their mature state, we developed a custom image analysis
software in MATLAB to detect the location of neurons, clusters of
neurons and neurite paths \cite{inpreparation}.
The performance of the algorithm is sketched in
Fig.~\ref{fig2}. The algorithm takes as an input a gray color image of the culture
at a particular day (Fig.~\ref{fig2}A), upon which it superimposes a
layer of new information comprising the contours of the clusters of neurons
(red shadows), the traces of the neurites (green lines), and
connection points between neurites, as well as those between neurites and clusters
(blue dots) (Fig.~\ref{fig2}B).

The information contained in the produced layer is then used to map the neuronal network into
a graph $\cal{G}$ (see Fig.~\ref{fig2}C) whose nodes (in blue) are
either cluster centroids or connection points, and the links (in
green) are straight lines connecting them. Therefore, our graph
is made of two types of nodes: neurons or clusters of neurons ($v_i$) and neurite
connection points ($u_i$). Treating all links as identical,
i.e. ignoring edge length and edge directionality, this graph can be described
in terms of a symmetric adjacency matrix $A$ whose elements $a_{ij}$
are equal to $1$ if nodes $i$ and $j$ are linked, and $0$ otherwise.

We focus on the network statistical properties
at the level of the $v_i$ nodes, ignoring the dynamics
of both neurite connections and branching points. Therefore, we extract
from $\cal{G}$ the subgraph defining the connectivity among nodes of
class $v_i$ in such a way that $v_i$ and $v_j$ are linked
either directly or through a connected path of $u_i$ nodes.

\begin{figure}
  \centering
\includegraphics[width=\textwidth]{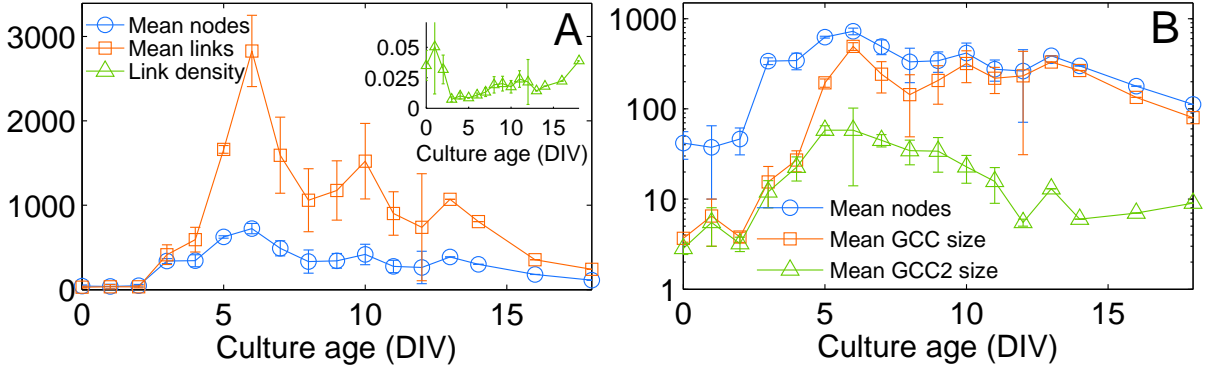}
   \caption{{\bf Density of the network as a function of culture
      age}. (A)  Mean number of nodes (blue circles),
    including neurons and clusters of neurons, and links
    connecting them (red squares), calculated for the 6 cultures vs.
    age (DIV).  Inset: the link  density (green triangles) quantifies the
    actual number of links divided by that of an all-to-all configuration
    [$N\cdot (N-1)/2$, being $N$ the number of connected
    nodes at each age]. (B) Log-linear plot of the mean number of nodes having
    at least one connection (blue circles), of the  mean size of the giant connected
    component (red squares) and of the second largest connected
    component (green triangles). In all plots, error bars stand for the standard errors of the mean (sem).
  }
\label{fig3}
\end{figure}

The analysis of the networks' evolution requires accounting for the birth and death
of links (and, in some cases, nodes) over time. Figure~\ref{fig3}
shows the mean values for the number of nodes and the of
links at each DIV, calculated for the 6 cultures.
During the growth phase, spanning from 0 to 6 DIV, the number of
nodes with at least one connection slowly increases with age,
while the number of links rises exponentially, reaching a
maximum at DIV 6. After this time point, the convergence of parallel neurites and
neuronal clusterization induces a more gentle decrease in the number
of links, accompanied by a slight reduction in the number of
nodes. In order to properly compare networks of different size,
we need to refer to a measure which is independent of the network
size: the link density, defined as
the ratio between the total number of measured links and the number of links
characterizing the arrangement of the same number of identified nodes
in a complete clique configuration. As illustrated in the inset of Fig.\ref{fig3}A, at any stage of
development, the cultured networks are far from being fully connected
(only about 2\% of all possible connections exist between nodes), and
thus operate in a low-cost regime of sparse anatomical connections.

In such a sparse connectivity regime, we quantify how our networks
constrained in 2D space percolate. To do so, we measure the size
$S_1$ of the giant connected component (GCC) and the
size $S_2$ of the second largest component (GCC2) as a function of the
age \cite{Bollobas}.
Figure~\ref{fig3}B shows that the number of nodes forming such connected components
smoothly increases at the same rate along the first days of the
network development, up to the DIV 6 when the difference in size between
them suddenly and consistently starts to grow.
From that point on, the
GCC2 starts collapsing and progressively merging into the GCC,
and the establishment of an almost fully connected network of clusters
characterizes the rest of the culture's life. Figure~\ref{fig3}B reports the evolution of the number
of nodes belonging to both GCC and GCC2. Although the network size is
too small to allow a general conclusion on how our two-dimensional spatial constraints are reflected in the percolation
properties, we can still estimate the percolation threshold $q_c$,
i.e. the fraction of connected nodes at the onset of the GCC2,
which is approximately $0.55$. This might indicate that our networks
belong to a different universality class of percolation, living in
between site percolation in a square lattice (that would correspond to
$q_c = 0.4$) and in a ER network (for which we would have
$q_c=1-1\langle k \rangle \sim 0.9$) \cite{Li2011}.

A deeper information on the culture evolution can be gathered
by monitoring the behavior of a subset of local and network-wide quantities.
For that purpose, we calculated several topological
properties of the extracted adjacency matrices (using the Matlab Boost
Graph Library package and the Brain
Connectivity Toolbox \cite{Rubinov2010}), whose definitions are provided in
\cite{PRBocca, Rubinov2010}. In particular, we analyzed the clustering coefficient ($C$),
the average shortest path length ($L$), the local ($E_{loc}$) and
global ($E_{glob}$) efficiency \cite{Latora2003}, the network
assortativity ($r$) and the cumulative degree distribution
($P_{cum}(k)$), obtained from the degree distribution $P(k)$ as $P_{cum}(k)=\sum_k^{k=k_{max}} P(k)$
being $k$ the degree (or number of links) of a node.

In all cases, the calculation of such statistics was restricted to
the set of nodes having at least one link,
and for the calculation of $L$  to those pairs of nodes
belonging to the GCC. Moreover, the experimental
values of $C$ and $L$ were also compared with those expected in
equivalent random null hypothesis networks, i.e. random networks artificially constructed
to have the same number of nodes and links and to display the same degree distribution. Specifically, for each
experimental network at a particular age, we generated 20 independent realizations of
 equivalent random networks, and calculated the corresponding expected
 network statistics.

Finally, in order to quantify the degree-degree correlation properties,
the network assortativity was defined by considering for each node
$i$ the average degree of its neighbors $k_{nn}$, and by computing the
linear regression of $\log(\langle k_{nn}\rangle)$ vs.
$\log(k_i^{p})$. The assortativity coefficient $r$ was then calculated as the Pearson
correlation coefficient corresponding to the best fit of $\log(\langle
k_{nn}\rangle)\sim p \log(k)$. If $r>0$ ($r<0$), the network is set to be
assortative (disassortative), while depending upon the obtained value of
$p$, the degree correlation properties are said to be of a linear ($p=1$), sub-linear ($p<1$), or super-linear ($p>1$) nature.

\section*{Results}
\subsection*{Emergence of small-world structure}

\begin{figure}
  \centering
\includegraphics[width=\textwidth]{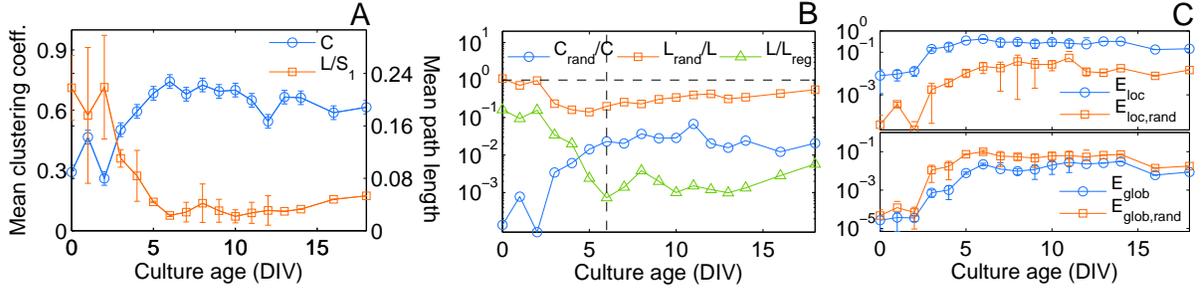}
  \caption{{\bf Network clustering and shortest path properties as a function of
    culture age}. (A) Absolute values of the clustering coefficient $C$ (blue circles, left axis)  and
    mean path length $L$ (red squares, right axis)  normalized to the size of the largest
    cluster. (B) Semi-log plot of normalized values of $C$ and
    $L$  with respect to the expected values for an equivalent random network having the same number of nodes and
    links and preserving the degree distribution: $C_{rand}/C$
    (blue circles) and $L_{rand}/L$ (red squares). The average
    path length is also compared to the value for a regular lattice as
    $L/L_{reg}$ (green   triangles) with $L_{reg}=S_1/\langle
    k\rangle$, being  $\langle k\rangle$ the average connectivity and
    $S_1$ the size of the largest connected component.
(C) Local (upper plot) and global (lower plot) efficiency as a
function of culture age and compared to their respective values for
the random graphs of the null model (see text for an explanation).
All quantities are averaged for the set of  6  cultures at each day of measure
    (DIV). As in the Caption of Fig.~\ref{fig3}, error bars represent the sem.
  }
\label{fig4}
  \end{figure}

The first days of the cultures' development (from
DIV 0 to DIV 3) were characterized by networks with very few nodes and
links (see Fig.~\ref{fig3}A). After DIV 3, the networks showed a very pronounced increase in the number
of links and nodes (from DIV 3 to DIV 6) preceding a spatial network reorganization
 eventually driving the graph into its clustered, mature state.

The associated networks statistics sheds light on the transition from
random to non-random properties with a progression of both the clustering
coefficient and the average path length (normalized by the GCC size) as a function of age (see Fig.~\ref{fig4}A).
The first significant result is the simultaneous increase in the
clustering coefficient and decrease in the mean path length,
a clear fingerprint of the emergence of a small-world network
configuration. This configuration becomes prominent at DIV 6 and stays
relatively stable through the last two weeks {\it in vitro}.
To properly asses the significance of this finding and isolate the
influence of the variable network size and
density, we calculated the values of $C$ and $L$  normalized
to the corresponding expected values for equivalent random (and lattice) null
model networks (see Fig.~\ref{fig4}B).
According to Watts and Strogatz's model \cite{WS98}, a small-world
network simultaneously exhibits short characteristic path length,
like random graphs, and high clustering, like regular lattices. Here,
we found a clear change in the trend at DIV 6 where $L_{rand}/L\le 1$, where
the average path length of the cultured network starts to be close to that of a random graph and
much smaller than that of a regular graph ($L_{reg}$ is calculated as
$L_{reg}=S_1/(2\langle k\rangle$)). At the same time, the clustering coefficient was
much higher (between 30-50 times) than that of the corresponding
random graphs.

These results are in agreement  with previous morphological
characterizations of {\it in vitro} neuronal networks at a single
developmental stage \cite{PREShefi}, where a similar small-world arrangement of connections was
evidenced at DIV 6.
However, to reinforce the evidence of the emergence of a small-world configuration
{\it during} the graph development
(as well as the fact that here the small-world metrics are not influenced
by network disconnectedness), we also measured the global and local
efficiency, as introduced by Latora and Marchiori in
\cite{Latora2003}.
These latter quantities, indeed, are seen
as alternative markers of the small-world phenomenon, in that
small-world networks are those propagating information efficiently
both at a global and at a local scale. The efficiency curves of the cultured
networks are reported in Fig.~\ref{fig4}C  as a function of age, and compared to the
efficiency of the equivalent random graphs. Our results indicate that
the connectivity structure of the neuronal networks evolve towards maximizing global efficiency
(making it similar to the value of random graphs), while promoting fault
tolerance by maximization of local efficiency (which is, instead, larger than the
local efficiency of a random graph), and both properties are realized at a relatively low
cost in terms of number of links (see again Fig.~\ref{fig3}A).

\subsection*{Node degree distribution evolution}

\begin{figure}
  \centering
\includegraphics[width=\textwidth]{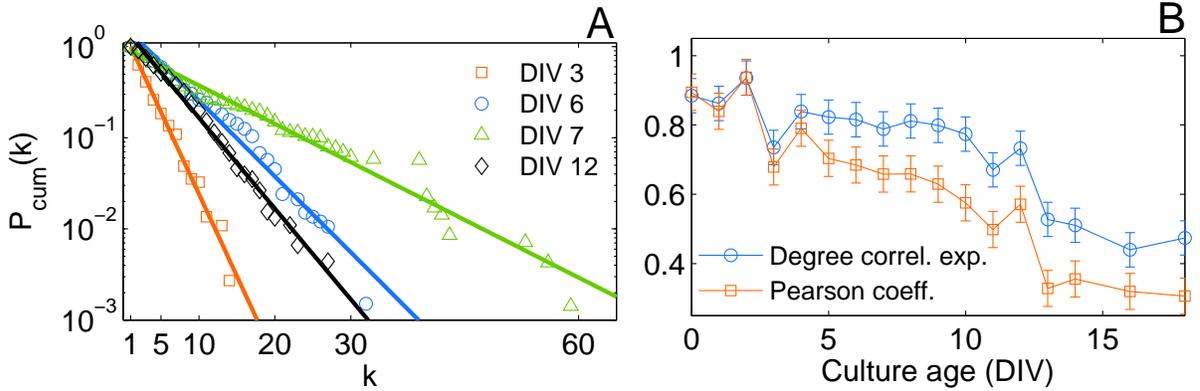}
  \caption{{\bf Degree distribution and degree-degree correlation.} (A) Cumulative node degree distributions on a semi-log
    scale for the state of the same culture at DIVs 3, 6, 7, and 12
    (see legend for the symbol coding).
    Solid lines correspond to the best exponential fitting $y(k)\sim
    \exp(- k/b)$, with $b\simeq \langle k\rangle$
  the mean degrees at DIV 3, 6, 7, and 12 respectively.
(B) Degree correlation exponent (blue
    circles) measuring the network assortativity and the corresponding
    Pearson coefficient (red squares). Both quantities are averaged for the set of  6
    cultures at each day of measure (DIV) and  error bars represent the sem.  }
  \label{fig5}
\end{figure}

Turning now our attention to network statistics at the micro-scale, we investigated how the node
degree distributions evolved during maturation process.
At all ages, cultures appeared to belong to the class of single-scale
networks, displaying a well defined characteristic mean node degree.
Figure~\ref{fig5}A shows that the cumulative degree
distributions $P_{cum}(k)$ for DIVs 3, 6, 7, and 12 had a fast decay
with a non monotonous increase in the average connectivity, with most
of the nodes having a similar number of connections and only a few
ones with degrees deviating significantly from such a number.

The data were fitted to an exponential scaling law $y(k)\sim exp
(-k/b)$ with very high goodness. The values of
the scaling parameter $b$ were close, within error, to the mean degree $\langle k\rangle$  of the networks at each culture age.
It has to be remarked that the distribution of node connections,
although always homogeneous, shifted during culture maturation toward
much broader distributions, with few highly connected
nodes appearing at DIVs 6 and 7.
These ``hubs'' at the peak days of the
culture evolution result from a branching process, allowing each
single neuron to reach a larger neighborhood.
Thus, at variance with scale-free networks \cite{RMPBarabasi,BarabasiSci99}, our cultured and clustered
networks are identified as a single-scale homogeneous population of
nodes. This is in agreement with reports on many other biological systems like the neuronal
network of the worm {\it   Caenorhabditis elegans} \cite{White1986,WD99}, and suggests the
existence of physical costs for the creation of new connections and/or
nodes limited capacity \cite{Amaral2000}.

While the number of neighbors (the degree) is a quantity retaining
information at the level of a single node, one can go further to
inspect degree-degree correlations, i.e. to quantify whether the
degrees of two connected nodes are correlated. It is known,
indeed, that biological networks feature generally disassortative
network structures \cite{Newman2002}, that is connections are more likely to be
established between high-degree and low-degree nodes. In our system,
we used the assortativity coefficient described in the Experimental
set-up section. Figure ~\ref{fig5}B shows the age evolution of
the Pearson coefficient $r$ and of the exponent $p$ of the degree correlation
($\langle k_{nn}\rangle = ak^p$). At one hand, as $r$ stays positive
during the whole development we can generally conclude that our networks are assortative
and, on the other hand, the trend of the degree correlation $p$
indicates that there is a transition from an almost linear (from DIV 0 to
DIV 2) to a sub-linear ($p\sim 0.7$) degree-degree correlation regime
during the small-world stage.

It is important to remark  here that, to the best of our knowledge,
this is the first report of assortativity in an {\it in vitro}
cultured neuronal network, and such an evidence actually
links to other studies  where assortativity was found in simple {\it
  in vivo} neuronal systems, like the {\it C. elegans} neural network structure \cite{Chatterjee2008}.

\subsection*{Spatial-growth model}

\begin{figure}
  \centering
\includegraphics[width=0.8\textwidth]{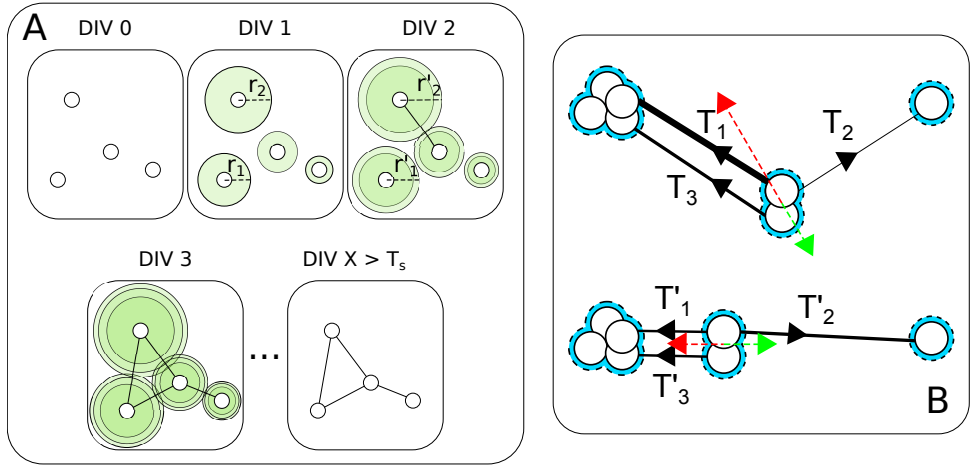}
  \caption{{\bf Growth model}. (A) Schematic representation of how
    cells get connected. At DIV 0, 4 cells of radius $a$ are located
    at random positions. The first iteration of the algorithm, DIV 1,
    assigns to each cell $i$ a disk of radius $r_i$ (green shade). At the next
    iteration, DIV 2, the disk's growth rate decreases, $r_i'$, and
    a link between two cells is established when their disks
    intersect (DIV 3). This process continues until $T_s$ steps. (B) Force diagram explaining
    cell migration and clustering. Tension forces $T_1$, $T_2$, and
    $T_3$ are acting on the central cluster composed of two
    cells, whose vector sum (red arrow) exceeds the adhesion to
  the substrate (green arrow). As a result, a new equilibrium
  state is produced with new tension forces $T_1'$,
  $T_2'$, and $T_3'$, being the central cluster pulled in the direction of the net
  force approaching the largest cluster.}
\label{fig6}
\end{figure}

A series of previous studies \cite{Shefi2002, Segev2003} singled out
tension along neurites and adhesion to the substrate as the two main
factors conditioning the neuronal self-organization into a clustered
network. Here we go a step ahead, and propose a simple spatial model
which not only incorporates migration of neurons but also explicitly
considers neurite growth, and the establishment of synaptic connections.

Our model is schematically illustrated in
Fig.~\ref{fig6}. We start by considering a set of $N$
cells. Each cell is a small disk of
radius $a$ randomly distributed in a 2-dimensional circular substrate of
area $S$. The algorithm then evolves the connections and positions of such disks at discrete times, each time step
$t$ corresponding to a DIV of the culture.
The complex process of neurite growth and the establishment of
synaptic connections is modeled by associating to each cell a time growing
disk in such a way that, two cells are linked at a given time step if their
outer rings intersect as shown for DIV's 2 and 3 in Fig.~\ref{fig6}A. At each time
step $t$, the radius $r_i\ge a$ increases by a quantity $\delta r_i$ which
decays as
 $$\delta r_i^t=\frac{V}{t} \left(1-\frac{1}{K_i}k_i^{t-1}\right)$$
\noindent
where $V$ is the neurite growth velocity (the same for all
cells), $K_i$ a random number in the interval $[1,N]$ and
$k_i^{t}$ the degree of the node (cell) at the time step $t$. The term $k_i^{t-1}/K_i$
introduces heterogeneity in the cell population,
and represents the fraction of links  acquired by the cell in the
previous steps from the initial randomly assigned endowment $K_i$. A very large $K_i$ indicates that,
potentially, a cell is very active and could connect many other
cells. The wiring process is iterated up to a given time step $T_s$, at which
the formation of new connections is stopped.

\begin{figure}
  \centering
 \includegraphics[width=\textwidth]{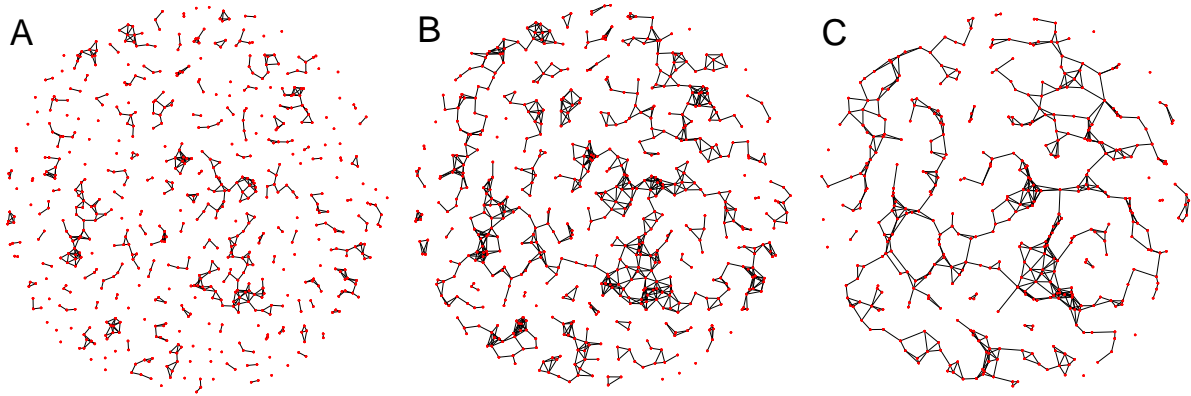}
  \caption{{\bf Schematic illustration of the network self-organization
    at three different instants of the automata generations} (A) DIV 3,
    (B) DIV 6, and (C) DIV 12. Simulation parameters: $N=700$, $V=42$, and $T_s=10$.  }
\label{fig7}
  \end{figure}

As for the process of cell migration and clusterization, cells or clusters whose distance is less than $2a$ are then
merged into the same new cluster. Furthermore, whenever two cells are
connected, an initial tension $\vec{T}_{ij}=0.1\vec{u}_{ij}$ is created
between them, and it is incremented in $0.1$ force units at each time
step, being $\vec{u}_{ij}$ the unit vector along the direction
connecting the two cells.
The total force acting on a cell or cluster $i$ is
given by $\vec{F}_i=\sum_j \vec{T}_{ij}$ with $j$ running over the
cell indexes connected to $i$, and not belonging to the same cluster.
Furthermore, each cell is ``anchored'' to the substrate by a
force $F_a=10$ force units, and the i$^{th}$ cell can only be detached if there is a net force $F_i$ acting on it larger than
$F_a$.  In the case of a cluster of cells, the adhesion force to the
substrate is considered to be the sum of the individual adhesions of the cells
composing the cluster. Therefore, cells and clusters move in a
certain direction in all circumstances in which the net force acting on them overcomes the adhesion force,
and an equilibrium point is
reached at a new position in which the new net force balances (or is
smaller) than the adhesion to the substrate (see Fig. \ref{fig6}B).

In order to validate our model, we ran a large number of simulations
for different values of the model parameters $N$, $V$ and $T_s$.
Remarkably, when comparing the statistical topological features of
the simulated networks to those measured from the set of
6 cultures, we found high correlation values exist only in a very narrow window
of $V$ and $T_s$. For instance, the parameter values which better fit
the experimental observations for $N=700$ are $V=40\pm 5$ and
$T_s=9\pm 1$.

Figure ~\ref{fig7} shows a typical output of the evolution of a simulated network.
The initial number of cells is taken to be of the same order
as in the experiments, and we chose as parameters $V$ and $T_s$ those with the
highest correlation with experiments. Boundary conditions mimic
the real experimental setup by canceling the adhesion force to the substrate
outside the culture area. The spatial organization of the network of
cells and clusters after 3, 6, and 12 DIV, closely resembles the one
observed in the experiments (see Fig.~\ref{fig1}).

\begin{figure}
  \centering
  \includegraphics[width=0.8\textwidth]{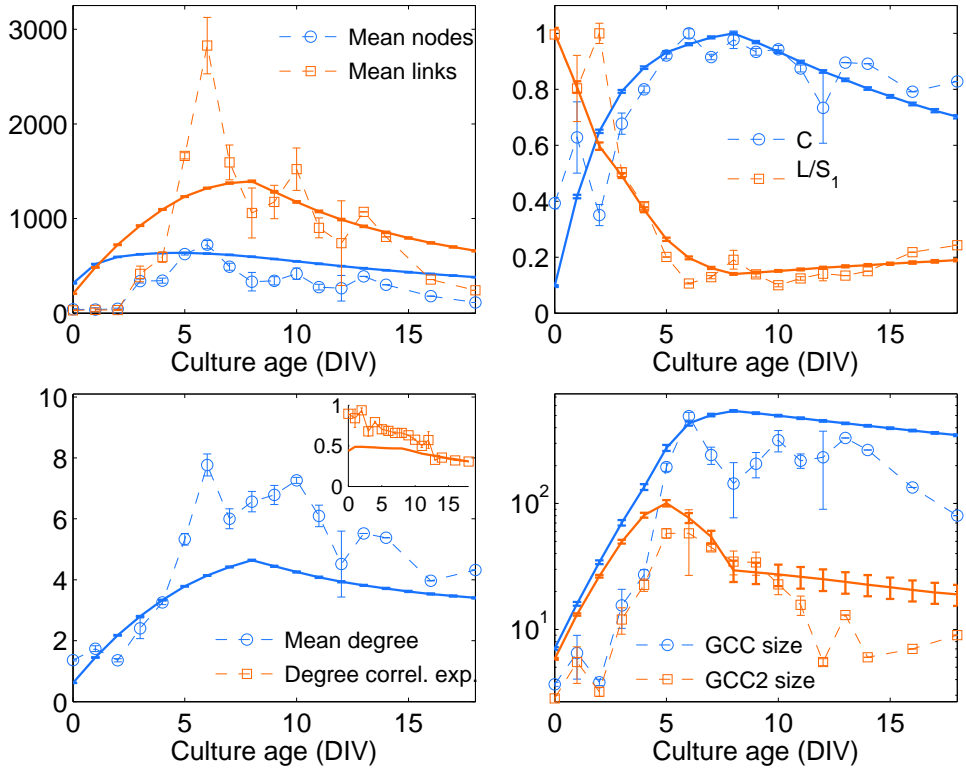}
  \caption{{\bf Comparison between model and experiment.} Legends in each panel clarifies on the topological
    quantities measured in experiments (dashed curves), and the corresponding trends of the
    simulated networks (solid curves). Simulation parameters are the same as in the
    caption of Fig.~\ref{fig7}, and each point is the ensemble average over 50 independent runs of the
    growth algorithm.}
\label{fig8}
  \end{figure}

Despite its relative simplicity, it is remarkable that the model offers
a rather good qualitative verification of the trends of
all the  structural network characteristics  measured in the experiments.
In particular, Fig.~\ref{fig8} reports a synoptic comparison of the mean number of nodes and
links, of $C$ and $L$, of the mean degree and degree correlation, and
of the sizes of the GCC and GCC2, measured in the experiments and
those obtained from the simulations of
the model with $N=700$, $V=42$, and $T_s=10$. The main observed difference
is found in the mean degree, the reason behind such a slight
discrepancy being that the model does not include any neurite branching
process, limiting the size of the neighborhood encompassed by the nodes.

Though it would have been unrealistic to expect a perfect quantitative
agreement between model and observations, the fact that
the model reproduces the main qualitative scenarios of the experiments
indicates that it captures the main processes underlying the observed
morphological evolution and self-organization of the cultures.

\section*{Discussion and conclusions}

In summary, we provided a large scale experimental investigation of the  morphological evolution of
{\it in vitro} primary cultures of dissociated invertebrate neurons
from locust ganglia. 
At all stages of the culture's development, we were able to identify
neurons' and neurites' location in automated way, and extract the adjacency matrix
that fully characterizes the connectivity structure of the networking neurons.
A systematic statistical analysis of a group of topological
observables has later allowed tracking of the main
network characteristics during the self-organization process of the
culture, and drawing important conclusions on the nature of the processes involved
in the culture' structuring. At early stages of development ($<$DIV 3)
characterized by a high neurite growth rate, homogeneous node degree distribution and low clustering resulted in a
random topology as expected given the fact that neurons were randomly
seeded. Following this immature period, neurite growth rate diminished
and tension along neurites started to shift the network  to a
small-world one with path lengths similar to random configurations but
presenting high clustering of connections. This transition from random
to small-world concurred with the percolation of the culture and the
onset of the giant connected network component. 

Furthermore, the identification of the main physical processes taking
place during the culture's morphological transformations, allowed us
to embed them into a simple growth model, qualitatively reproducing the
overall scenario observed in the experiments.

Our results extend previous studies where network
properties of cultures were investigated at a particular developmental
stage  and for a lesser number of nodes \cite{PREShefi}. These results also
systematically characterize several topological network measures
along the entire culture's evolution, and unveil many yet unknown self-organization properties,
such as {\it i}) the fact that a small-world configuration spontaneously emerges
in connection to a large increase in both local and global network's efficiency, 
and {\it ii}) the evidence that cultures tend to organize in a regime of non trivial degree mixing which,
in turn, is characterized by assortative degree-degree
correlation features. The evolution from an initial random to
a small-world topology has also been reported recently in the context
of a functional network of a cortical culture
\cite{Downes2012}. However, although functional connectivity
correlates well with anatomical connectivity, there are studies
showing that strong functional connections may exist between nodes
with no direct physical connection \cite{Honey2009}. This suggests that future studies
are needed in which both anatomical and functional networks are
accessible in order to understand their complex entanglement. 

Given the absence of external chemical or electrical stimulations,
we conclude that such complex network evolution and morphological structuring is indeed
an intrinsic property of neuronal maturation. Our study therefore contributes to the understanding of the complex
processes ruling the morphological structuring
of cultured neuronal networks as they self-organize from collections of separated cells into clustered graphs, and may help
identifying culture development stages in new, specific and targeted, experiments.

\section*{Acknowledgments}
The authors acknowledge financial support from the Spanish Ministerio de Ciencia e Innovaci\'on (Spain) under
project FIS2009-07072, and of Comunidad de Madrid
(Spain) under project MODELICO-CM S2009ESP-1691. DSS is supported by
the Comunidad de Madrid through the R+D Activity Program NEUROTEC-CM (2010/BMD-2460).

\bibliography{biblio}

\begin{thebibliography}{10}
\bibitem{inpreparation}
The used imaging software has been fully customized for the purpose of the
  analysis of the present data. The general details of the developed imaging
  software will be reported elsewhere.
\bibAnnoteFile{inpreparation}
\end{thebibliography}


\end{document}